\documentclass[3p,times,twocolumn]{elsarticle}
 \biboptions{comma,sort&compress}
 
\usepackage{graphicx}
\usepackage{amsmath}
\usepackage{here}
\usepackage{ecrc}

\volume{00}

\firstpage{1}

\journalname{Nuclear and Particle Physics Proceedings}
\runauth{Stephan Narison}

\jid{nppp}
\jnltitlelogo{Nuclear and Particle Physics Proceedings}

\usepackage{amssymb}

\usepackage[figuresright]{rotating}

\begin{document}

\begin{frontmatter}

\title{
%
Tau data-driven evaluation of the Hadronic Vacuum Polarization\,$^*$} 
 
 \cortext[cor0]{Talk given by Álex Miranda at the 26th International Conference in Quantum Chromodynamics (QCD23), 10-14 July 2023, Montpellier - FR.}

 \author[label1,label2]{Pere Masjuan
}
   \address[label1]{Grup de F\'isica Te\`orica, Departament de F\'isica, Universitat Aut\`onoma de Barcelona, 08193
Bellaterra (Barcelona), Spain.
}
\ead{masjuan@ifae.es}

  \author[label2]{Alejandro Miranda
}
   \address[label2]{Institut de F\'isica d’Altes Energies (IFAE) and The Barcelona Institute of Science and
Technology (BIST), Campus UAB, 08193 Bellaterra (Barcelona), Spain.
}
\ead{jmiranda@ifae.es}

  \author[label3]{Pablo Roig
}
   \address[label3]{Departamento de F\'isica, Centro de Investigaci\'on y de Estudios Avanzados del Instituto
Polit\'ecnico Nacional, Apdo. Postal 14-740,07000 Ciudad de M\'exico, M\'exico
}
\ead{pablo.roig@cinvestav.mx}

\pagestyle{myheadings}
\markright{ }
\begin{abstract}
\noindent
Windows in Euclidean time have become a standard tool for comparing lattice QCD and data-driven computations of the hadronic vacuum polarization (HVP) contribution to the muon $g-2$. Here we review our results, obtained using isospin-rotated $\tau^-\to\pi^-\pi^0\nu_\tau$ data
 instead of $e^+e^-\to\pi^+\pi^-$ measurements, and compare them to other approaches. Consistency of the tau-based and lattice results hints to underestimated uncertainties in the $e^+e^-$ data. If that is the case, the theory prediction of the muon $g-2$ would only lie at $\lesssim2.5\sigma$ from its measured value.
\end{abstract}
\end{frontmatter}
\section{Introduction}
The measurement of the anomalous magnetic moment of the muon, $a_\mu=(g_\mu-2)/2$, is becoming more and more precise, thanks to the recent FNAL data \cite{Muong-2:2021ojo, Muong-2:2023cdq} and the legacy BNL result \cite{Muong-2:2006rrc}, all agreeing remarkably. The experimental world average is 
\begin{equation}\label{eq.amuExp}
a_\mu^{\mathrm{Exp}}=116592059(22)\times10^{-11}\,.
\end{equation}

The corresponding Standard Model (SM) prediction, at a similar accuracy, is much more challenging, as we briefly review in the following.

The QED \cite{Aoyama:2012wk,Aoyama:2019ryr} and Electroweak \cite{Czarnecki:2002nt,Gnendiger:2013pva} contributions are known with uncertainties negligible compared to that in Eq.~(\ref{eq.amuExp}). The problem is on the hadronic pieces, specifically on the dominant HVP one, with an error that -according to the White Paper (WP) \cite{Aoyama:2020ynm}- basically doubles the experimental one, Eq.~(\ref{eq.amuExp})~\footnote{We will not cover here the other hadron contribution, given by the hadronic light-by-light piece, with an uncertainty smaller than $a_\mu^{\mathrm{Exp}}$. Its SM prediction in the WP \cite{Aoyama:2020ynm} is based on Refs.~\cite{Melnikov:2003xd,Masjuan:2017tvw,Colangelo:2017fiz,Hoferichter:2018kwz,Gerardin:2019vio,Bijnens:2019ghy,Colangelo:2019uex,Pauk:2014rta,Danilkin:2016hnh,Jegerlehner:2017gek,Knecht:2018sci,Eichmann:2019bqf,Roig:2019reh,Colangelo:2014qya}, with a precision matched by the most recent lattice QCD computations \cite{Chao:2021tvp,Blum:2023vlm}. Later developments, mainly for the most difficult contributions, coming from axial-vector mesons and remaining short-distance constraints, are covered in Refs.~\cite{Hoferichter:2020lap,Knecht:2020xyr,Masjuan:2020jsf,Ludtke:2020moa,Bijnens:2020xnl,Bijnens:2021jqo,Zanke:2021wiq,Colangelo:2021nkr,Leutgeb:2021bpo, Miranda:2021lhb,Hoferichter:2021wyj,Miramontes:2021exi,Bijnens:2022itw,Radzhabov:2023odj,Ludtke:2023hvz, Hoferichter:2023tgp}. See our recent account \cite{PP}.}.

Traditionally, $a_\mu^{\mathrm{HVP}}$ was obtained from $\sigma(e^+e^-\to\mathrm{hadrons}$), via a dispersive integral with a kernel peaked at low energies, as the cross-section itself is (apart from resonances) \cite{Aoyama:2020ynm}. This makes that $\sim 73\%$ of $a_\mu^{\mathrm{HVP}}$ comes from the $\pi^+\pi^-$ contribution~\cite{Aoyama:2020ynm,Davier:2017zfy,Keshavarzi:2018mgv,Colangelo:2018mtw,Hoferichter:2019mqg,Davier:2019can,Keshavarzi:2019abf}, with $\sim80\%$ of the overall uncertainty stemming from this channel. Then, the disagreement between the $e^+e^-$ data-driven $a_\mu^{\mathrm{SM}}$ prediction and $a_\mu^{\mathrm{Exp}}$ (always $a_\mu^{\mathrm{Exp}}>a_\mu^{\mathrm{SM}}$) comes mostly from the $\pi^+\pi^-$ contribution.
In this channel, the long-standing discrepancy between BaBar \cite{BaBar:2009wpw,BaBar:2012bdw} and KLOE \cite{KLOE:2004lnj,KLOE:2008fmq,KLOE:2010qei,KLOE:2012anl,KLOE-2:2017fda} data was still marginally acceptable for the White Paper combination \cite{Aoyama:2020ynm} (yielding a $5.0\sigma$ discrepancy with $a_\mu^{\mathrm{Exp}}$), but the recent CMD-3 measurement \cite{CMD-3:2023alj,CMD-3:2023rfe} has decreased the overall compatibility (see also Refs.~\cite{CMD-2:2006gxt,BESIII:2015equ,SND:2020nwa}). CMD-3 alone would give $a_\mu^{\mathrm{SM}}$ less than one $\sigma$ away from $a_\mu^{\mathrm{Exp}}$.

There is only one lattice QCD computation of $a_\mu^{\mathrm{HVP}}$ with competitive precision to the data-based evaluations, the one reported by the BMW collaboration~\cite{Borsanyi:2020mff}, which agrees at less than $2\sigma$ with $a_\mu^{\mathrm{Exp}}$.

In this unclear situation for $a_\mu^{\mathrm{SM}}$, it is worth recalling that an alternative data-driven evaluation is possible, replacing $e^+e^-\to\pi^+\pi^-$ by isospin-rotated $\tau^-\to\pi^-\pi^0\nu_\tau$ data, as was pioneered in Ref.~\cite{Alemany:1997tn} and later on pursued in several other analyses \cite{Narison:2001jt,Cirigliano:2001er,Cirigliano:2002pv,Davier:2002dy,Davier:2003pw,Maltman:2005yk,Maltman:2005qq,Davier:2010fmf,Davier:2010nc,Benayoun:2012etq,Davier:2013sfa,Narison:2023srj,Esparza-Arellano:2023dps}, including Ref.~\cite{Miranda:2020wdg}, which computed the isospin-breaking corrections (see also \cite{Escribano:2023seb}) our work \cite{Masjuan:2023qsp} reported here is mainly based upon.

\section{Method and results}

The leading order (LO) contribution to $a_\mu^{\mathrm{HVP}}$ is traditionally calculated as
\begin{equation}\label{eq.amuHVPLO}
a_\mu^{\mathrm{HVP,LO}}=\frac{1}{4\pi^3}\int_{s_\text{thr}}^{\infty}\mathrm{d}s\, K(s) \,\sigma^0_{e^+e^-\to\mathrm{hadrons }(\gamma)}(s)\,,
\end{equation}
where the upper-index zero signals that the bare cross-section~\footnote{$\sigma^0$ is obtained from the dressed cross section by applying mass-dependent corrections for vacuum polarization and by adding back the effects of final-state radiation (which should belong to $a_\mu^{\mathrm{HVP,NLO}}$ but are included for convenience in Eq.~(\ref{eq.amuHVPLO}) instead).} is used.

An alternative data-driven evaluation is viable, replacing $\sigma^0(e^+e^-\to\pi^+\pi^-)$ with the spectrum of the $\tau^-\to\pi^-\pi^0\nu_\tau$ decays $\big(\mathrm{d}\Gamma_{\tau^-\to\pi^-\pi^0\nu_\tau(\gamma)}/\mathrm{d}s\big)$
~\cite{Belle:2008xpe,ALEPH:2005qgp,CLEO:1999dln,OPAL:1998rrm}. In this way,
\begin{equation}\label{eq.pipixsection}
\sigma^0_{e^+e^-\to\mathrm{hadrons }(\gamma)}(s)=\left[\frac{K_\sigma(s)}{K_\Gamma(s)}\frac{\mathrm{d}\Gamma_{\tau^-\to\pi^-\pi^0\nu_\tau(\gamma)}}{\mathrm{d}s}\frac{R_\text{IB}(s)}{S_\text{EW}}\right]\,,
\end{equation}
where the ratio of $K$ functions depends on the kinematics and absorbs global constants, and $S_\text{EW}$ is the universal short-distance electroweak correction factor~\cite{Marciano:1988vm}, $S_\text{EW}=1.0233(3)$ \cite{Alemany:1997tn}. The remaining isospin-breaking corrections are encapsulated in \cite{Cirigliano:2001er,Davier:2010fmf}
\begin{equation}\label{eq.RIB}
 R_{\text{IB}}(s)=\frac{\text{FSR}(s)}{G_{\text{EM}}(s)}\frac{\beta^3_{\pi^+\pi^-}}{\beta^3_{\pi^-\pi^0}}\Bigg|\frac{F_V(s)}{f_+(s)}\Bigg|^2.
\end{equation}
Two contributions to $R_\text{IB}(s)$ are easy to evaluate: the ratio of $\beta$ functions and the final-state radiation term, $\text{FSR}(s)$. A non-negligible model dependence is presently associated to the neutral-to-charged current form factor ratio $\left(\frac{F_V(s)}{f_+(s)}\right)$ and to the long-distance electromagnetic corrections for the di-pion tau decays ($G_\text{EM}(s)$).

The effect of the $S_\text{EW}$ correction on $a_\mu$ is $-119.6\times10^{-11}$, while the phase space correction yields $-74.7\times10^{-11}$, both with negligible uncertainties. The FSR effect amounts to $+45.6(4.6)\times10^{-11}$ \cite{Miranda:2020wdg}, in agreement with \cite{Davier:2010fmf}.

The ratio of the form factors is challenging and depends on the different $\rho$ masses and widths according to their charge, as well as on the $\rho-\omega$ mixing present only in the neutral channel. In Ref.~\cite{Miranda:2020wdg} we followed both the proposals of Refs.~\cite{Cirigliano:2002pv,Davier:2010fmf} to compute this correction, resulting in the contributions $+77.8(24.0)\times10^{-11}$ and $+41.0(48.9)\times10^{-11}$, compatible with these references. This is currently the dominant error of the tau-based prediction of $a_\mu^{\mathrm{HVP,\pi\pi}}$ and could be reduced with improved measurements of the $\rho^{\pm,0}$ pole positions and of the $\Gamma(\rho\to\pi\pi\gamma)$ channels.

We use the  $G_\text{EM}(s)$ correction computed \cite{Miranda:2020wdg} in Chiral Perturbation Theory \cite{Weinberg:1978kz,Gasser:1983yg,Gasser:1984gg} with resonances \cite{Ecker:1988te,Ecker:1989yg}~\footnote{This framework was also used to compute contributions to the hadronic light-by-light piece of $a_\mu$ in Refs.~\cite{Roig:2014uja,Guevara:2018rhj,Roig:2019reh}.}, as first done in Ref.~\cite{Cirigliano:2002pv}. We have:\\
- Included the operators of the original Lagrangian \cite{Ecker:1988te,Ecker:1989yg}, as in Ref.~\cite{Cirigliano:2002pv}, and first evaluated its associated uncertainty including the subleading terms in the chiral expansion \cite{Cirigliano:2006hb,Kampf:2011ty} that are fixed by QCD short-distance constraints~\cite{Cirigliano:2006hb,Kampf:2011ty,Roig:2013baa}. We will name this approach $\mathcal{O}(p^4)$ (since it includes all operators that -upon resonances integration- contribute to the chiral low-energy constants at this chiral order) from now on, and yields our reference results.\\
- Included additional operators, which are suppressed at low energies \cite{Cirigliano:2006hb,Kampf:2011ty} and estimated those unrestricted by perturbative QCD plus phenomenology by chiral counting~\cite{Cirigliano:2006hb,Kampf:2011ty,Roig:2013baa}. In this last case there are so many free couplings that the uncertainties are artificially large, with a shift in the central value that seems to be overestimated. This approximation will be called $\mathcal{O}(p^6)$ in the following and is given for completeness.\\
Our results for the $G_\text{EM}(s)$ correction to $a_\mu$ in these two cases are $\left(-15.9_{-16.0}^{+\,\,5.7}\right)\times10^{-11}$ (consistent with Refs.~\cite{Cirigliano:2002pv,Flores-Baez:2006yiq,Davier:2010fmf}) and $\left(-76\pm46\right)\times10^{-11}$, respectively.

\begin{figure}
    \centering
    \includegraphics[width=7.7cm]{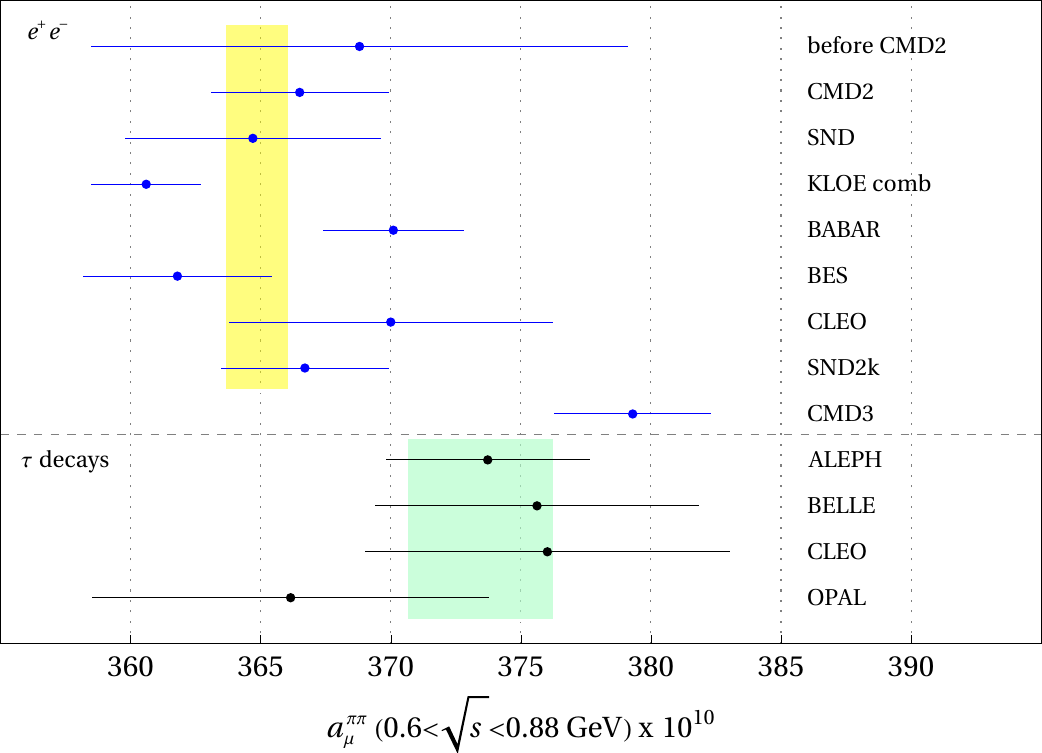}
    \caption{The $\pi\pi(\gamma)$ contribution to the $a_\mu^{\text{HVP,LO}}$ in the energy range $0.6<\sqrt{s}<0.88\text{ GeV}$ obtained from $e^+e^-\to\pi^+\pi^-$ cross section and di-pion $\tau$ decays.}
    \label{fig:amu_eetau}
\end{figure}

Fig.~\ref{fig:amu_eetau} displays the di-pion contribution to $a_\mu^\text{HVP, LO}$ in the $\rho$ resonance region obtained using either $\sigma(e^+e^-\to\mathrm{hadrons})$ (top part of the plot, with mean in yellow) or the $\tau^-\to\pi^-\pi^0\nu_\tau$ spectrum (bottom of the plot, with mean in green). A larger value of $\sim10\times10^{-10}$ is obtained with tau data. The CMD-3 data point is a clear outlier among the $e^+e^-$ results (in excellent agreement with the tau-based ones) and so it was not averaged with the rest.

The tau-based evaluation of $a_\mu^{\text{HVP, LO}}$ yields $\left(704.1^{+4.1}_{-4.0}\right)\times10^{-10}$ ($\left(699.1^{+6.1}_{-5.2}\right)\times10^{-10}$ including subleading operators) \cite{Miranda:2020wdg}~\footnote{We note that these results have shifted by $-1.6\times10^{-10}$, since we corrected our double counting of the subleading non-logarithmic short-distance correction for quarks, as noted in Ref.~\cite{Davier:2023fpl} (see Ref.~\cite{Cirigliano:2023fnz}).}, in good agreement with the BMW Collaboration lattice QCD result ($(707.5\pm5.5)\times10^{-10}$) and with $a_\mu^{\mathrm{Exp}}$ (as well as with the CMD-3-based prediction). It is then interesting to scrutinize further this accord in different energy regions or, as it has become conventional in the lattice computations, using different windows in Euclidean time.

For this, we will employ the weight functions in center-of-mass energy $\tilde{\Theta}(s)$ \cite{Colangelo:2022vok}, related to those in Euclidean time by~\cite{RBC:2018dos}
\begin{eqnarray}\label{eq:weight_function}
    \Theta_{SD}(t)&=&1-\Theta(t,t_0,\Delta),\\
    \Theta_{win}(t)&=&\Theta(t,t_0,\Delta)-\Theta(t,t_1,\Delta),\\
    \Theta_{LD}(t)&=&\Theta(t,t_1,\Delta),\\
    \Theta(t,t^\prime,\Delta)&=&\frac{1}{2}\left(1+\tanh{\frac{t-t^\prime}{\Delta}}\right),
\end{eqnarray}
defining the short-distance ($SD$), intermediate ($win$) and long-distance ($LD$) windows ($t_0=0.4\text{ fm}$, $t_1=1.0\text{ fm}$, $\Delta=0.15\text{ fm}$). We note that $LD$ dominates up to $\sqrt{s}\sim0.9\text{ GeV}$, $SD$ from $\sqrt{s}\sim2.3\text{ GeV}$ on, and $win$ in between (we will also use $int$ for this one), see Fig. 1 in Ref.~\cite{Colangelo:2022vok}.

\begin{figure}
    \centering
    \includegraphics[width=7.7cm]{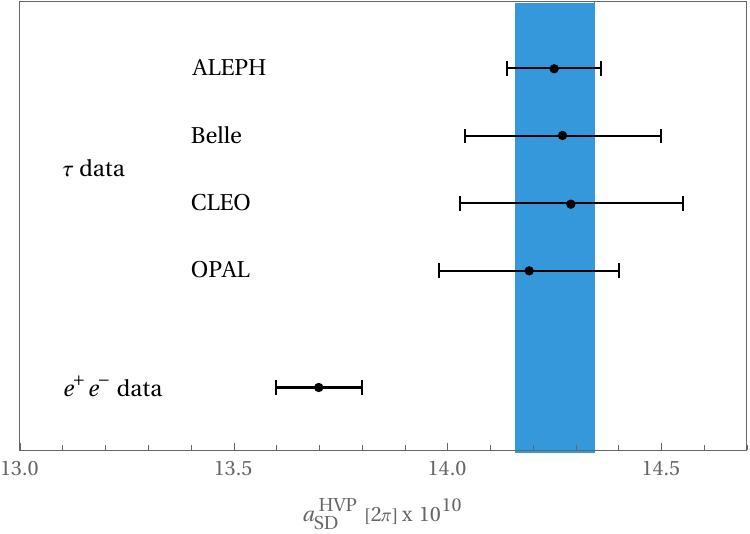}
    \includegraphics[width=7.6cm]{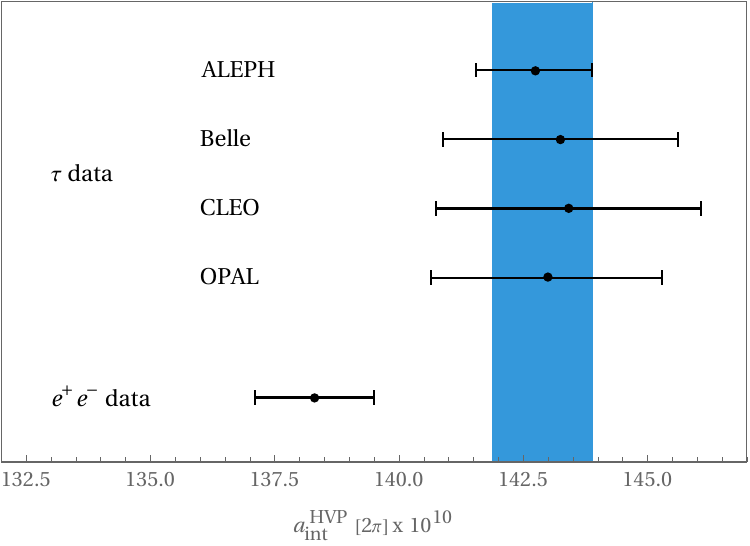}
    \includegraphics[width=7.7cm]{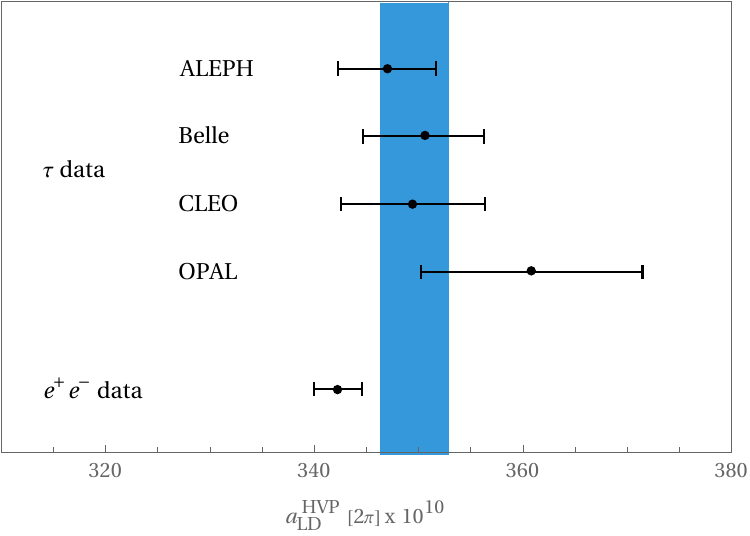}
    \caption{Window quantities ($SD$ top, $int$ medium and $LD$ bottom) for the $2\pi$ contribution below $1.0\,\text{GeV}$ to $a_\mu^{\mathrm{HVP}}$ at $\mathcal{O}(p^4)$. 
 The blue region shows the experimental average from $\tau$ data. The $e^+e^-$ number is taken from Ref. \cite{Colangelo:2022vok}. }
    \label{fig:ChPTOp4}
\end{figure}

\begin{figure}
    \centering
    \includegraphics[width=7.8cm]{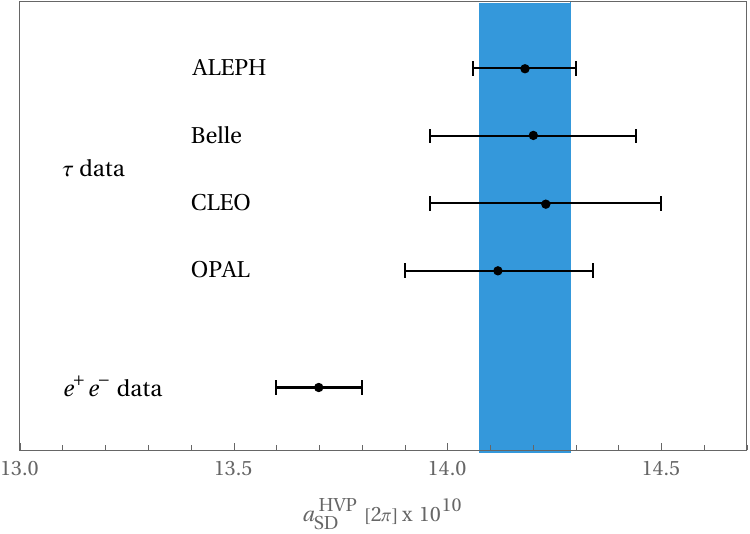}
    \includegraphics[width=7.7cm]{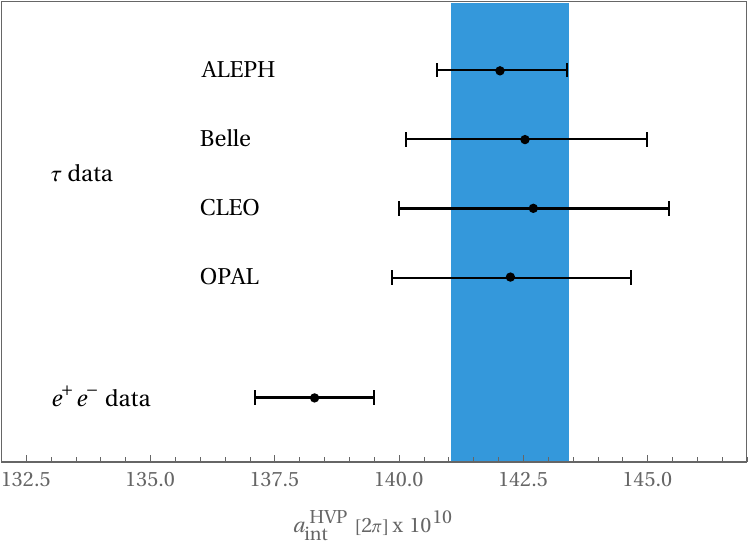}
    \includegraphics[width=7.8cm]{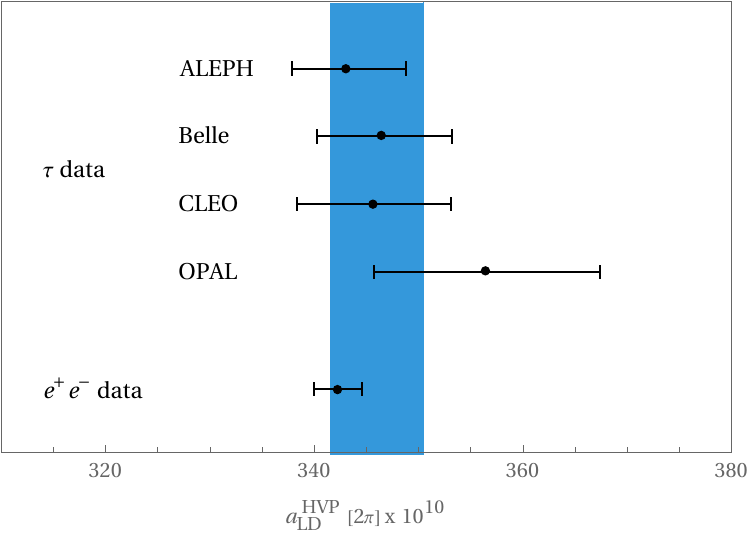}
    \caption{Analog to Fig. \ref{fig:ChPTOp4} but at $\mathcal{O}(p^6)$.}
    \label{fig:ChPTOp6}
\end{figure}

We have evaluated the di-pion tau-based \cite{Belle:2008xpe,ALEPH:2005qgp,CLEO:1999dln,OPAL:1998rrm} contribution to $a_\mu^{\mathrm{HVP, LO}}$ using the windows explained above, with the quoted values of $t_{0,1}$ and $\Delta$. In Ref.~\cite{Masjuan:2023qsp} we provide tables separating the corrections from each source of isospin-breaking in Eqs.~(\ref{eq.pipixsection}) and (\ref{eq.RIB}) and showing the results for every experiment separately. We plot the results for the three different window contributions to $a_\mu^{\mathrm{HVP}}$ (note that they scale as $\sim1:10:25$) in Figs.~\ref{fig:ChPTOp4} and \ref{fig:ChPTOp6}. These graphs nicely display the consistency among the different tau measurements. In the $SD$ and $int$ windows, $e^+e^-$ data-based results (from Ref.~\cite{Colangelo:2022vok}) and tau values disagree markedly. Only in the last plot of Fig. \ref{fig:ChPTOp6} (for the $LD$ window), the big errors on the $G_\text{EM}$ correction yield compatibility at one $\sigma$ between both (which is not the case for the reference $\mathcal{O}(p^4)$ result in the last plot of Fig. \ref{fig:ChPTOp4}).

\begin{figure}
    \centering
    \includegraphics[width=8cm]{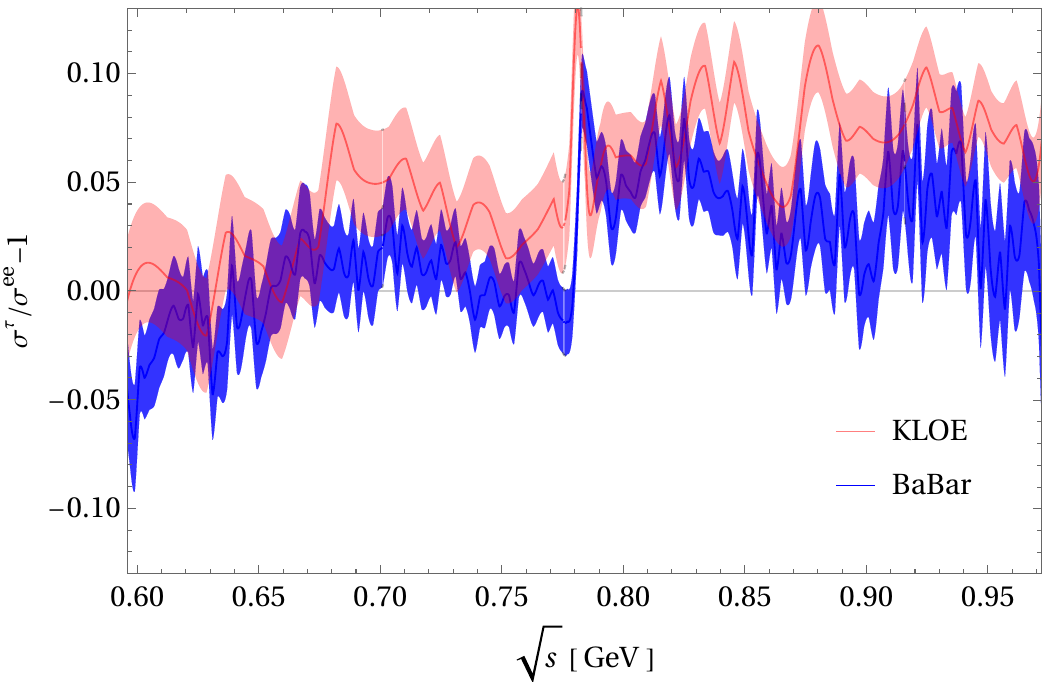}
    \includegraphics[width=8cm]{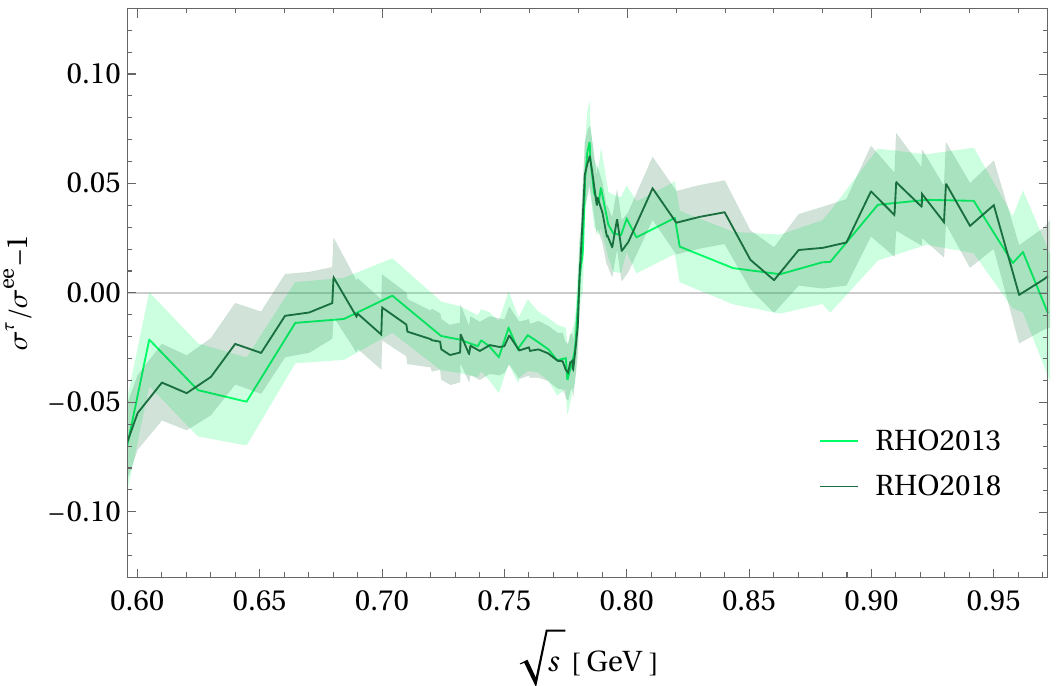}   
    \caption{Comparison between the $\tau$ (after IB corrections) and the $e^+e^-\to\pi^+\pi^-$ spectral functions using the ISR measurements from BABAR~\cite{BaBar:2012bdw} and KLOE~\cite{KLOE:2012anl} (top) and the energy-scan measurements from CMD-3~\cite{CMD-3:2023alj} (bottom). 
}
    \label{fig:KLEO_BABAR}
\end{figure}

Fig. \ref{fig:KLEO_BABAR} magnifies the comparison between the IB-corrected $\tau^-\to\pi^-\pi^0\nu_\tau$ and the $e^+e^-\to\pi^+\pi^-$ spectral functions using the ISR measurements from BABAR~\cite{BaBar:2012bdw} and KLOE~\cite{KLOE:2012anl} (top panel) and the energy-scan measurements from CMD-3~\cite{CMD-3:2023alj} (bottom panel). Colored bands show the weighted average of the uncertainties coming from the data sets in each figure. Although it may seem that an enhanced form factors' ratio correction could improve agreement between tau and $e^+e^-$ CMD-3 data (see e.g. figs. 16 and 17 in ref.~\cite{Miranda:2020wdg}), further studies are needed to fully understand this (even more so in the comparison with BaBar and KLOE data).

We cannot contrast directly our tau-based results with the lattice outcomes. For this we need to supplement ours with $e^+e^-$ data where needed. We have done this in two ways, to estimate the corresponding uncertainty. First,  we have subtracted the contribution from the $2\pi$ channel below $1.0\,\text{ GeV}$ to the values reported in Table 1 of Ref.~\cite{Colangelo:2022vok} (`$<1$ GeV' stands for this procedure) and replaced it by our corresponding mean values. Second, we have rescaled the contributions from the $2\pi$ channel using the full evaluation of $a_{\mu}^{\text{HVP, LO}}[\pi\pi,e^+e^-]$ in Refs.~\cite{Davier:2019can,Keshavarzi:2019abf}, removed it from the total contribution and substituted it by our values (without adding any symbol to represent this proceeding).

\begin{figure}
    \centering
    \includegraphics[width=8cm]{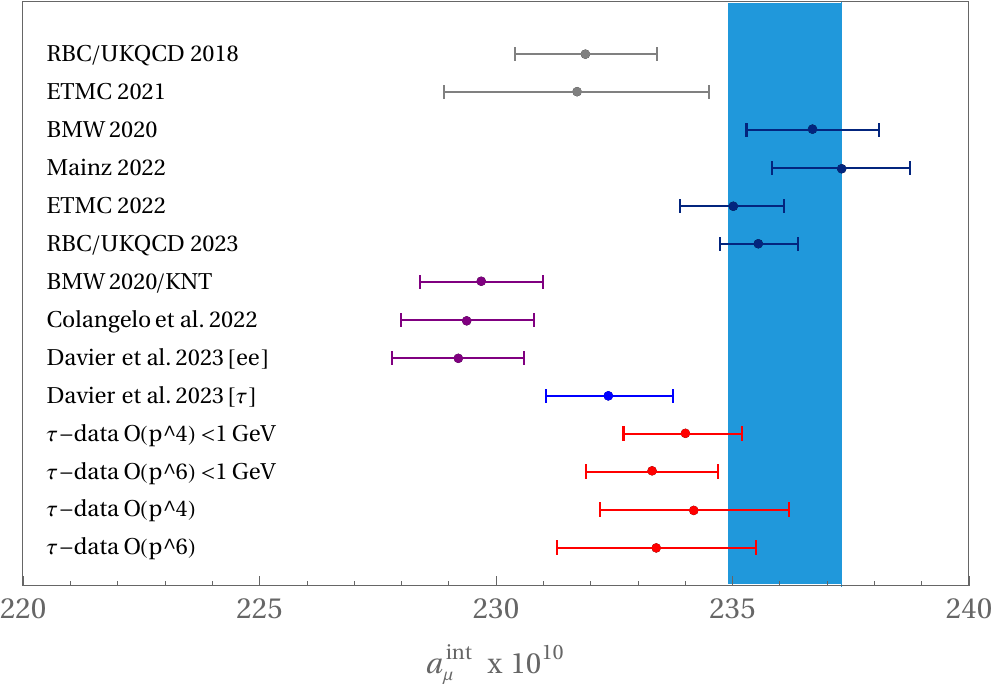}
    \caption{Comparison of the total intermediate window contribution to $a_{\mu}^{\text{HVP, LO}}$ according to lattice QCD, $e^+e^-$ and $\tau$ data-driven evaluations. The blue band corresponds to the weighted average of the lattice results excluding RBC/UKQCD 2018~\cite{RBC:2018dos} and ETMC 2021~\cite{Giusti:2021dvd}.}
    \label{fig:lattice_results}
\end{figure}

Our results are displayed in Fig. \ref{fig:lattice_results} for the intermediate window, where the blue band shows the weighted average of the lattice results, $a_\mu^{int}=235.8(6)\cdot 10^{-10}$, excluding those from RBC/UKQCD 2018~\cite{RBC:2018dos} and ETMC 2021~\cite{Giusti:2021dvd} collaborations. Tau-data based  contributions in the intermediate window are significantly closer to the lattice QCD values than to the $e^+e^-$ ones. Thus, the $\sim 4.3\sigma$ discrepancy between the $e^+e^-$ data-driven and lattice evaluations shrinks to $\sim 1.2\sigma$ using $\tau$ data for the $2\pi$ channel. There is only one lattice result for the short-distance window~\cite{ExtendedTwistedMass:2022jpw} which seems to agree with both data-driven HVP evaluations (although more closely with the tau-based). See Table \ref{HVP:tab4.4c} and Fig. \ref{fig:lattice_results}.

\begin{table}[htbp]
\centering
\resizebox{8cm}{!}{\begin{tabular}{|c|c|c|c|c|}
\hline
 \multicolumn{5}{ |c| }{$a_\mu^{\text{HVP,LO}}$} \\ [0.3ex]
\hline
 & SD & int & LD & Total\\ [0.3ex]
\hline
$\tau$-data $\mathcal{O}(p^4)$ $\leq 1\text{ GeV}$ & $69.0(5)$ & $234.0(^{1.2}_{1.3})$ & $402.5(^{3.3}_{3.4})$ & $705.5(^{5.0}_{5.2})$ \\ [0.3ex]
$\tau$-data $\mathcal{O}(p^6)$ $\leq 1\text{ GeV}$ & $68.9(5)$ & $233.3(1.4)$ & $398.5(^{4.9}_{4.2})$ & $700.7(^{6.8}_{6.1})$ \\ [0.3ex]
\hline
$\tau$-data $\mathcal{O}(p^4)$ & $69.0(7)$ & $234.2(2.0)$ & $402.6(^{3.8}_{3.9})$ & $705.8(^{6.5}_{6.6})$\\ [0.3ex]
$\tau$-data $\mathcal{O}(p^6)$ & $68.9(7)$ & $233.4(2.1)$ & $398.5(^{5.3}_{4.6})$ & $700.8(^{8.1}_{7.4})$\\ [0.3ex]
\hline
RBC/UKQCD 2018~\cite{RBC:2018dos}  & $-$ & $231.9(1.5)$ & $-$ & $715.4(18.7)$\\ [0.3ex]
ETMC 2021~\cite{Giusti:2021dvd}  & $-$ & $231.7(2.8)$ & $-$ & $-$ \\ [0.3ex]
BMW 2020~\cite{Borsanyi:2020mff}  & $-$ & $236.7(1.4)$ & $-$ & $707.5(5.5)$\\ [0.3ex]
Mainz/CLS 2022~\cite{Ce:2022kxy}  & $-$ & $237.30(1.46)$ & $-$ & $-$\\ [0.3ex]
ETMC 2022~\cite{ExtendedTwistedMass:2022jpw}  & $69.33(29)$ & $235.0(1.1)$ & $-$ & $-$ \\ [0.3ex]
RBC/UKQCD 2023~\cite{Blum:2023qou}  & $-$ & $235.56(82)$ & $-$ & $-$\\ [0.3ex]
\hline
WP~\cite{Aoyama:2020ynm} & $-$ & $-$ & $-$ & $693.1(4.0)$\\ [0.3ex]
BMW 2020/KNT~\cite{Keshavarzi:2018mgv,Borsanyi:2020mff}  & $-$ & $229.7(1.3)$ & $-$ & $-$ \\ [0.3ex]
Colangelo et al. 2022~\cite{Colangelo:2022vok} & $68.4(5)$ & $229.4(1.4)$ & $395.1(2.4)$ & $693.0(3.9)$\\ [0.3ex]
Davier et al. 2023 [$e^+e^-$]~\cite{Davier:2023cyp} & $-$ & $229.2(1.4)$ & $-$ & $694.0(4.0)$\\ [0.3ex]
Davier et al. 2023 [$\tau$]~\cite{Davier:2023fpl} & $-$ & $232.4(1.3)$ & $-$ & $-$\\ [0.3ex]
\hline
\end{tabular}}
\caption{Window quantities for $a_\mu^{\text{HVP,LO}}$ in units of $10^{-10}$. The first and second pairs of rows differ in the way tau data is complemented with $e^+e^-$ measurements (as explained in the main text). 
The rows 5-10 are the lattice results~\cite{RBC:2018dos,Giusti:2021dvd,Borsanyi:2020mff,Ce:2022kxy,ExtendedTwistedMass:2022jpw,Blum:2023qou}. The last three rows are the evaluations obtained using $e^+e^-$ data, and the WP number \cite{Aoyama:2020ynm} is shown for reference. See Fig. 11 in Ref.~\cite{Blum:2023qou} for more details.}
\label{HVP:tab4.4c}
\end{table}

\section{Conclusions}
There is a puzzle in $e^+e^-\to\pi^+\pi^-$ data (between CMD-3, KLOE and BaBar, mostly) that currently limits the precision for $a_\mu^{\mathrm{HVP,LO}}$ and thus for the whole $a_\mu^\text{SM}$ in this data-driven way. At the moment there is only one lattice QCD evaluation (by BMW) at a competitive precision in the whole range, which agrees with CMD-3 and $a_\mu^{\mathrm{Exp}}$.

In order to understand this situation, comparisons between lattice QCD and $e^+e^-$-based results for $a_\mu^{\mathrm{HVP}}$ have become standard by using windows in Euclidean time.

We recall that alternative data-driven evaluations are possible and worth, using di-pion tau decay data instead of the corresponding $e^+e^-$ measurements (accounting for the required isospin-breaking corrections, including structure-dependence). These were extremely useful before the very precise KLOE and BaBar measurements and have traditionally been closer to $a_\mu^{\mathrm{Exp}}$. After reviewing our tau-based analysis \cite{Miranda:2020wdg} for the $\pi\pi$ contribution to $a_\mu^{\mathrm{HVP,LO}}$, we summarize our recent work \cite{Masjuan:2023qsp}. We verify that our agreement with lattice results extends from the whole integrated effect to the three considered windows, which reinforces agreement of $a_\mu^{\mathrm{SM}}$ with $a_\mu^{\mathrm{Exp}}$ at $\lesssim2.5\sigma$. Further work seems needed to reach compatibility between $e^+e^-\to\pi^+\pi^-$ measurements by the different experiments. Our results can also be valuable for lattice efforts \cite{Bruno:2018ono} addressing the computation of the relevant IB-corrections needed to use di-pion tau decay data as illustrated here.

\section*{Acknowledgements}
A. Miranda thanks the organizing committee of QCD23 for this interesting conference. P. M. has 
been supported by the European Union’s Horizon 2020 Research and Innovation Programme under grant 824093 (H2020-INFRAIA- 2018-1), the Ministerio de Ciencia e Innovación under grant
PID2020-112965GB-I00, and by the Secretaria d’Universitats i Recerca del Departament d’Empresa
i Coneixement de la Generalitat de Catalunya under grant 2021 SGR 00649. IFAE is partially
funded by the CERCA program of the Generalitat de Catalunya. A. M. is also supported by 
MICINN with funding from European Union NextGenerationEU (PRTR-C17.I1) and by Generalitat de Catalunya. P. R. is funded by Conahcyt and Cinvestav.
\vfill\eject

\end{document}